\documentclass[trackchanges]{aastex701}
\usepackage{amsmath}

\begin{document}

\title{An Enhanced Formation Channel for Galactic Dual-Line Gravitational-Wave Sources: von Zeipel-Lidov-Kozai Effect in Triples Involving Sgr A*}

\author[orcid=0000-0001-5033-6168, gname=Wen-Fan, sname=Feng]{Wen-Fan Feng}
\affiliation{Kavli Institute for Astronomy and Astrophysics, Peking University,
Beijing 100871, China}
\email[show]{fengwf@pku.edu.cn}  

\author[orcid=0000-0002-2898-1360, gname=Tan, sname=Liu]{Tan Liu}
\affiliation{School of Fundamental Physics and Mathematical Sciences, Hangzhou Institute for Advanced Study, University of Chinese Academy of Sciences, Hangzhou 310024, China}
\affiliation{School of Physical Sciences, University of Chinese Academy of Sciences, Beijing 100049, China}
\email[]{lewton@mail.ustc.edu.cn}

\author[orcid=0000-0003-0065-8622, gname=Yun, sname=Fang]{Yun Fang}
\affiliation{Institute of Fundamental Physics and Quantum Technology, Ningbo University, Ningbo 315211, China}
\affiliation{School of Physical Science and Technology, Ningbo University, Ningbo 315211, China}
\email[]{fangyun@nbu.edu.cn}

\author[orcid=0000-0001-7402-4927, gname=Yacheng, sname=Kang]{Yacheng Kang}
\affiliation{Department of Astronomy, School of Physics, Peking University, Beijing 100871, China}
\affiliation{Kavli Institute for Astronomy and Astrophysics, Peking University, Beijing 100871, China}
\email[]{yckang@stu.pku.edu.cn}

\author[orcid=0000-0002-0643-8295, gname=Bin, sname=Liu]{Bin Liu}
\affiliation{Institute for Astronomy, School of Physics, Zhejiang University, Hangzhou 310027, China}
\email[]{liubin23@zju.edu.cn} 

\author[orcid=0000-0002-1334-8853, gname=Lijing, sname=Shao]{Lijing Shao}
\affiliation{Kavli Institute for Astronomy and Astrophysics, Peking University,
Beijing 100871, China}
\affiliation{National Astronomical Observatories, Chinese Academy of Sciences,
Beijing 100101, China}
\email[show]{lshao@pku.edu.cn}

\begin{abstract}

The dense Galactic Center environment is expected to host compact binary inspirals detectable by future space-borne gravitational wave (GW) observatories (e.g., LISA, TianQin, Taiji) in the millihertz band. Aided by information from these facilities, next-generation ground-based GW detectors (e.g., Cosmic Explorer, Einstein Telescope) can potentially capture gravitational radiation in the hectohertz band from rapidly spinning neutron star (NS) components in such binaries. These Galactic Center systems are thus anticipated to act as dual-line (i.e., low-frequency inspiral and high-frequency spin) GW sources. However, the formation channels of these systems remain largely unexplored. 
In this \textit{Letter}, we propose that the von Zeipel-Lidov-Kozai (ZLK) effect can enhance the formation of dual-line GW sources in hierarchical triples involving the Galactic supermassive black hole, Sgr A*. We show that ZLK-driven oscillations in the eccentricity and inclination of the inner binary can modulate the GW emission from both the binary inspiral and the individual NS spins. This effect boosts the expected dual-line source count by a factor of $\sim 5\text{–}10$, from rare to $\mathcal{O}(1)$ in 4 years, making dual-line observations substantially more probable. Our results demonstrate that the ZLK effect may provide an important formation channel for Galactic dual-line GW sources. 

\end{abstract}



\section{Introduction}

The groundbreaking detection of gravitational waves (GWs) has opened an unprecedented window for observing the cosmos \citep{LIGOScientific:2016aoc, LIGOScientific:2017vwq}. With their distinct, high-sensitivity frequency bands, next-generation ground-based detectors (e.g., Cosmic Explorer, \citet{CE2022slt}; Einstein Telescope, \citet{ET2010}) and space-borne observatories (e.g., LISA, \citet{LISA2017}; TianQin, \citet{TianQin2016}; Taiji, \citet{Taiji2017}) will facilitate the detection of low-frequency inspiral and high-frequency spin (dual-line\footnote{Dual-line originally refers to GW emission at $2f_{\rm orb}$ and $2f_{\rm s}$ from circular-orbit inspiralling NS binaries \citep{Tauris:2018kzq}, where $f_{\rm orb}$ and $f_{\rm s}$ denote the orbital frequency and NS spin frequency, respectively. For eccentric-orbit binaries and precessing NS components, both the low- and high-frequency GW emission exhibit multiple harmonics; we nevertheless retain this terminology for simplicity.}) gravitational radiation from Milky Way neutron star (NS) binaries, including NS--NS, NS--black hole (NS--BH), and NS--white dwarf (NS--WD) systems. This dual-line signal comprises two components: millihertz-band radiation from the binary inspiral, and hectohertz-band radiation from a rapidly spinning NS component. 
Population simulation and GW waveform modeling of isolated NS--NS systems in the Galactic field, treated as dual-line sources, have been performed in previous works \citep{Feng:2025jnx, Feng:2024ect, Feng:2023gpe}. Observations of such systems would enable constraints on NS structural parameters \citep{Tauris:2018kzq, Chen:2021mrf, Suvorov:2021mhr, Feng:2024ulg, Feng:2025jnx}, thereby helping to pin down the long-sought equation of state of NS matter.
Critically, dual-line GW detection uniquely breaks the intrinsic degeneracy between NS structural parameters and source distance, a fundamental limitation unresolvable with spinning NS GW observations alone. Binary inspiral emission delivers independent distance calibration, resolving this degeneracy to enable robust NS property measurements and tight constraints via NS spin precession \citep{Feng:2025jnx}. 

The Galactic Center hosts Sagittarius A$^*$ (Sgr A$^*$), the nearest known supermassive black hole (SMBH), with a mass of $\sim 4\times 10^6~{\rm M}_\odot$ \citep{Ghez:2003qj, Ghez:2008ms, Gillessen:2008qv, Boehle:2016imz, EventHorizonTelescope:2022wkp}. Dominating the gravitational dynamics of its surroundings, Sgr A$^*$ provides a unique laboratory for testing gravity and stellar dynamics \citep[e.g.,][]{Ghez:2003rs, Alexander:2005jz, Gillessen:2011aa, Hopman:2009gz, Alexander:2013ewq, Hees:2017aal, Chu:2017gvt, Shao:2018klg, EventHorizonTelescope:2022xqj, Hu:2024blq, Yu:2025apk}.
The dense environment at the Galactic Center likely harbors abundant stellar and compact binaries, many of which form triple systems with Sgr A$^*$. Stability constraints favor a hierarchical configuration for most such systems: a tight inner binary orbited by a distant tertiary on a wider orbit, forming the outer binary. Gravitational perturbations from the distant tertiary (e.g., Sgr A$^*$) drive periodic oscillations in the inner binary's orbital eccentricity and inclination via the von Zeipel-Lidov-Kozai effect (ZLK; \citealp{VonZeipel1910, Kozai:1962zz, Lidov:1962wjn}; see \citealp{Naoz:2016cjb} for a review). This effect induces rich and complex dynamical processes in stellar evolution within the Galactic Center \citep[e.g.,][]{Stephan:2016kwj}. 
Studies of the ZLK effect have concentrated on stellar-mass BH-BH systems in galactic nuclei. The prospects for detecting ZLK-modulated GW signals with LISA have been assessed \citep{Hoang:2019kye, Randall:2019sab, Knee:2024mst, Grishin:2025ofa}, while analytical frameworks for GW modeling from ZLK-driven inspiraling systems have been formulated \citep{Chandramouli:2021kts}.

Despite large uncertainties in parameter settings and binary evolution, \citet{Stephan:2019fhf} modeled the dynamical evolution of binaries near Sgr A$^*$ under the ZLK effect, including tidal interactions, relativistic effects, and single/binary stellar evolution. Their simulations produced diverse outcomes, including compact binaries detectable as GW sources by LISA and Advanced LIGO \citep{KAGRA:2013rdx}.
Incorporating ultra-stripped supernova explosions with reduced natal kicks, \citet{Wang:2020jsx} estimated that, neglecting ZLK-driven eccentricity oscillations for post-evolutionary compact binaries, LISA could detect 0.4--4 NS--NS and 0.2--2 NS--BH systems within the Galactic Center's inner parsec, corresponding to merger rates of $\sim 0.3~\mathrm{Gpc^{-3}~yr^{-1}}$. These estimates are conservative, as they exclude additional dynamical processes like binary–single/binary–binary interactions \citep[e.g.,][]{Rodriguez:2011ef, Zhang:2019puc, Sedda:2020wzl} and single–single captures \citep[e.g.,][]{OLeary:2008myb, Tsang:2013mca, Hoang:2020gsi} that could enhance the compact binary population.
For dual-line sources, assuming log-uniform outer-orbit distributions and a detection fraction of $20\%$--$60\%$ for rapidly spinning NS components \citep{Feng:2025jnx}, we conservatively predict only 0.01--0.4 detectable systems within 100 AU over 4 years of combined observations. This suggests that the Galactic Center dual-line detections remain extremely challenging and unlikely with current expectations.

However, this pessimistic outlook is significantly altered when the ZLK effect is taken into account. In this \textit{Letter}, we investigate dual-line GW sources from NS--NS systems in hierarchical triples involving Sgr A$^*$, demonstrating that the ZLK effect can enhance detection rates by a factor of $\sim 5\text{–}10$.
Our dynamical analysis shows that ZLK oscillations dominate over general relativistic precession for typical NS--NS systems at the Galactic Center. We specifically model the GWs from the spinning NS components subject to the ZLK effect and map the detectable parameter space (equatorial ellipticity vs. spin period) accessible to Cosmic Explorer. The ZLK-induced eccentricity excitation facilitates dual-line signal formation, transforming detection count from a conservative estimate of 0.01--0.4 to 0.05--4 over 4 years and making dual-line observation in dense environments observationally promising.
Throughout this paper, we adopt geometric units ($G=c=1$).

\section{Dynamical timescales for Galactic Center triples}
\label{sec-timescales}

\begin{figure}[t]
\centering
\includegraphics[scale=0.7]{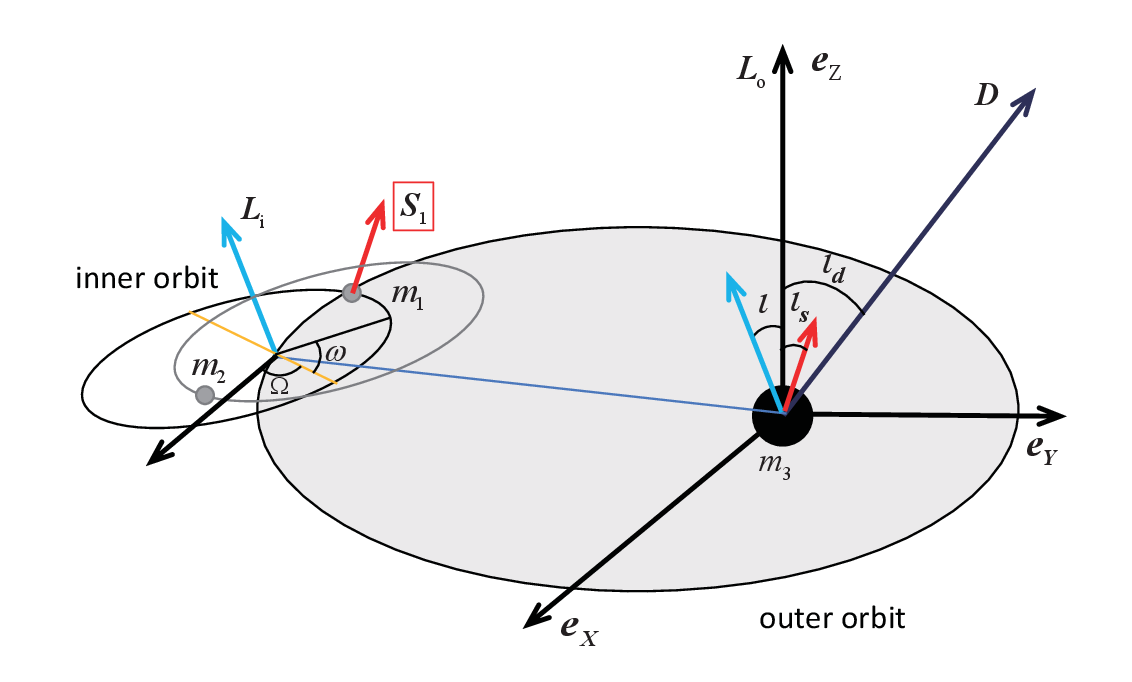}
\caption{Geometric configuration of the hierarchical triple system where an NS binary orbits Sgr A$^*$. The reference frame is chosen such that the outer orbital angular momentum $\boldsymbol{L}_{\rm o}$ aligns with the $Z$-axis. The inner binary orbital plane is tilted at inclination $\iota$ with respect to the outer orbit, with longitude of ascending node $\Omega$ and pericenter angle $\omega$ defining the orbital orientation. The vector $\boldsymbol{S}_1$ represents the primary NS spin. The detector inclination angle $\iota_{d}$ is defined between $\boldsymbol{L}_{\rm o}$ and $\boldsymbol{D}$, and the spin inclination angle $\iota_{s}$ between $\boldsymbol{L}_{\rm o}$ and $\boldsymbol{S}_1$. Initially, both $\boldsymbol{D}$ and $\boldsymbol{S}_1$ lie in the $Y$-$Z$ plane, and their $Y$-components are positive.}
\label{fig:orbitgeometry}
\end{figure}

We consider a hierarchical triple system in the Galactic Center (see Fig.~\ref{fig:orbitgeometry}), consisting of an inner binary orbiting the Sgr A$^*$ of mass $m_3$. As a representative case, we model the inner binary as an NS--NS system with a spinning primary component (mass $m_1$, spin $\boldsymbol{S}_1$) and a non-spinning companion (mass $m_2$) \citep{Feng:2023fez}. Our methodology extends straightforwardly to NS--BH systems. The inner and outer orbits are described by their respective semimajor axes ($a_{\rm i}, a_{\rm o}$) and eccentricities ($e_{\rm i}, e_{\rm o}$).
The angular momenta of the inner and outer orbits are denoted by $\boldsymbol{L}_{\rm i}$ and $\boldsymbol{L}_{\rm o}$, with their respective magnitudes $L_{\rm i}$ and $L_{\rm o}$.
The outer orbital plane is adopted as the reference plane, such that the outer orbital angular momentum is aligned with the $Z$-axis. The validity of this reference frame is justified by the dynamical timescale analysis presented below. 
The inclination angle $\iota_{d}$ is measured between $\boldsymbol{L}_{\rm o}$ and the vector towards detector $\boldsymbol{D}$, while $\iota_{s}$ is the angle between $\boldsymbol{L}_{\rm o}$ and primary spin $\boldsymbol{S}_1$. The reference system is defined with $\boldsymbol{D}$ projected onto the outer orbital plane as the $Y$-axis and the $X$-axis determined by the right-hand rule, such that $\boldsymbol{S}_1$ initially resides in the $Y$-$Z$ plane and its $Y$-component is positive.
Further details on the definitions of the inner orbital elements are provided in \citet{poisson2014gravity}.

Following the Galactic Center binary population simulation results presented by \citet{Wang:2020jsx}, the system parameters adopted in our analysis are set to $m_1 = m_2 = 1.4~{\rm M}_\odot$, $m_3 = 4 \times 10^6~{\rm M}_\odot$, $a_{\rm i}=0.012~{\rm AU}$,  $e_{\rm i}=0.6$, $a_{\rm o}=2500~m_3$ (corresponding to a physical semimajor axis of $\simeq 100~{\rm AU}$), and $e_{\rm o}=0.9$.
The orbital periods of the inner and outer orbits are
\begin{align}
P_{\rm i} &= 7~{\rm h}~\left( \frac{2.8~{\rm M}_{\odot}}{m_1+m_2} \right)^{1/2} \left( \frac{a_{\rm i}}{0.012~{\rm AU}} \right)^{3/2} , \\
P_{\rm o} &= 0.5~{\rm yr}~\left( \frac{m_3}{4\times 10^6~{\rm M}_{\odot}} \right) \left( \frac{a_{\rm o}}{2500~m_3} \right)^{3/2} .
\end{align}

Sgr A$^*$ can perturb the inner orbit via the ZLK effect, in which oscillations in eccentricity and inclination arise on a timescale \citep{Liu:2019tqr}:
\begin{align}
P_{\rm{ZLK}} &\sim 20~{\rm yr}~\left(\frac{1 - e_{\rm o}^2}{1 - 0.9^2} \right)^{3/2} 
\left( \frac{m_3}{4\times 10^6~{\rm M}_{\odot}} \right)^2 
\left( \frac{a_{\rm o}}{2500~m_3} \right)^3 \left( \frac{m_1 + m_2}{2.8~{\rm M}_\odot} \right)^{1/2} 
\left( \frac{a_{\rm i}}{0.012~\rm{AU}} \right)^{-3/2} .
\end{align}

The inner binary undergoes periastron precession at 1st post-Newtonian (PN) order due to general relativity, with a precession period given by \citep{robertson1938note}:
\begin{equation}\label{eq-1pn}
P_{\rm 1PN} \sim 220~{\rm yr}~\left(\frac{1 - e_{\rm i}^2}{1 - 0.6^2} \right) \left( \frac{m_1 + m_2}{2.8~{\rm M}_\odot} \right)^{-3/2} \left( \frac{a_{\rm i}}{0.012~{\rm AU}} \right)^{5/2} .
\end{equation}

The NS spin $\boldsymbol{S}_1$ is coupled to the inner orbital angular momentum $\boldsymbol{L}_{\rm i}$ through de-Sitter precession (1.5PN effect; e.g., \citealt{Barker:1975ae}) with a period
\begin{equation}
P_{S_1 L_{\rm i}} \sim 260~{\rm yr}~\left(\frac{1 - e_{\rm i}^2}{1 - 0.6^2} \right)  \left( \frac{a_{\rm i}}{0.012~{\rm AU}} \right)^{5/2} .
\end{equation}

The inner $\boldsymbol{L}_{\rm i}$ will undergoes a de Sitter-like precession around the outer $\boldsymbol{L}_{\rm o}$ \citep[e.g.,][]{Liu:2019tqr, Yu:2020dlm} with a period 
\begin{equation}
P_{L_{\rm i} L_{\rm o}} \sim 160~{\rm yr}~\left(\frac{1 - e_{\rm o}^2}{1 - 0.9^2} \right) \left( \frac{m_3}{4\times 10^6~{\rm M}_{\odot}} \right) \left( \frac{a_{\rm o}}{2500~m_3} \right)^{5/2} .
\end{equation}

If Sgr A$^*$ possesses a spin angular momentum $\boldsymbol{S}_3$ with magnitude $S_3=\chi_3 m_3^2$ (where $\chi_3$ is the dimensionless spin parameter), then $\boldsymbol{L}_{\rm o}$ undergoes precession around $\boldsymbol{S}_3$ with a period \citep[e.g.,][]{Liu:2021uam, Fang_2019}
\begin{align}
P_{L_{\rm o} S_3} &\sim 2.8\times 10^3~{\rm yr}~\left(\frac{1 - e_{\rm o}^2}{1 - 0.9^2} \right)^{3/2} 
\left( \frac{0.9}{\chi_3} \right) 
\left(  \frac{m_3}{4\times 10^6~{\rm M}_{\odot}} \right) \left( \frac{a_{\rm o}}{2500~m_3} \right)^3 .
\end{align}
The precession frequency of $\boldsymbol{L}_{\rm i}$ around $\boldsymbol{S}_3$ is one-fourth that of $\boldsymbol{L}_{\rm o}$ around $\boldsymbol{S}_3$, thus the corresponding timescale is much longer \citep{Laeuger:2023qyz}.

The time it takes for the NS--NS to merge due to GW radiation can be written as \citep{Peters:1964zz}
\begin{equation}
\tau_{\rm gw}= 1.2\times 10^8~{\rm yr}~ \frac{F(0.6)}{F(e_{\rm i})}  \left( \frac{\mathcal{M}}{1.2~{\rm M}_\odot} \right)^{-5/3} \left( \frac{P_{\rm i}}{7~\rm{h}} \right)^{8/3} ,
\end{equation}
with the eccentric orbital radiation factor $F(e)={\left( 1+\frac{73}{24}{{e}^{2}}+\frac{37}{96}{{e}^{4}} \right)}/{{{\left( 1-{{e}^{2}} \right)}^{7/2}}}$ and the inner binary chirp mass $\mathcal{M}=(m_1 m_2)^{3/5}/(m_1 + m_2)^{1/5}$. A large gravitational radiation decay timescale ensures the monochromaticity of harmonic frequencies.

Our analysis shows that, for the typical wide-orbit inner NS binaries taken from previous triple system simulations \citep{Wang:2020jsx}, ZLK timescales are much shorter than general relativistic precession timescales, establishing ZLK oscillations as the dominant dynamical mechanism.

\section{Detectability of neutron stars in Galactic Center triples}
\label{sec-NSwavesInTriple}

The ZLK effect drives periodic oscillations in the orbital eccentricity and inclination of the compact inner NS binary, directly modulating its dual-line GW radiation. Previous investigations have concentrated on binary inspiral signatures, leaving continuous NS radiation significantly underexplored. Here, we address this gap by modeling individual NS gravitational radiation under the ZLK effect at the Galactic Center. 

As a proof of concept, we adopt a widely used model for NS gravitational radiation \citep[e.g.,][]{Maggiore:2007ulw}, in which the NS is asymmetric and rotates rapidly about one of its own principal axes of inertia, with the GW frequency being twice the rotation frequency. Our calculation can be easily extended to the precessing triaxial NSs \citep{Feng:2024ect}.
Under the quadrupole approximation, the two GW polarizations in the source frame are given by
\begin{subequations} \label{eq-2polarizations}
\begin{align}
h_{+}(t) &= \frac{4 \, I_3 \, \epsilon \, \Omega_{\rm r}^2 \, \bigl[ 1 + \cos^2(\iota_d - \iota_s) \bigr] \cos( 2 \, t \, \Omega_{\rm r} )}{2D} , \\
h_{\times}(t) &= \frac{4 \, I_3 \, \epsilon \, \Omega_{\rm r}^2 \, \cos( \iota_d - \iota_s) \sin(2 \, t \, \Omega_{\rm r})}{D} ,
\end{align}
\end{subequations}
where $D$ is the distance between the source and the detector, $\Omega_{\rm r}$ is the rotation frequency of the NS (not to confuse with angle $\Omega$), $\epsilon$ is the equatorial ellipticity given by $(I_2-I_1)/I_3$, and $I_3$ is the moment of inertia of the NS with respect to the principal axis aligned with the rotation axis, while the other two moments of inertia $I_1$ and $I_2$ are perpendicular to it, $\iota_d$ and $\iota_s$ denote the inclination angles of the detector and the NS spin relative to the reference-frame $Z$-axis, respectively.

A time-dependent GW strain at the detector is given by $h(t)=F_{+}(t)h_{+}(t)+F_{\times}(t)h_{\times}(t)$, where $F_{+}(t)$ and $F_{\times}(t)$ are the antenna pattern functions \citep{Jaranowski:1998qm}. Due to the motion of Earth-based detectors relative to the Solar System Barycenter (SSB) and that of the NS about the triple system’s barycenter, the observed continuous wave signal undergoes Doppler modulation from these two effects. As such, the GW angular frequency measured at the detector can be expressed as \citep[e.g.,][]{Feng:2023gpe, Feng:2024ect, Covas:2019jqa}:
\begin{equation}\label{eq:Dopshift}
\Omega_{\rm r}^{\rm d} \approx  \Omega_{\rm r} \left(1+ {\boldsymbol n}_{\rm d}  \cdot \frac{{\rm d}\boldsymbol{r}_{\rm ns}}{{\rm d}t} + {\boldsymbol n}  \cdot \frac{{\rm d}\boldsymbol{r}_{\rm d}}{{\rm d}t} \right) .
\end{equation}
Here ${\boldsymbol n}$ and $\boldsymbol{r}_{\rm d}$ are the triple source direction vector and the detector position vector in the SSB frame \citep{Jaranowski:1998qm},
$-{\boldsymbol n}_{\rm d}=(0,\sin{\iota}_{\rm d},\cos{\iota}_{\rm d})$ is the SSB position vector in the triple source frame. The position vector $\boldsymbol{r}_{\rm ns}$ of the NS with spin $\boldsymbol{S}_1$ in this frame is given by
%
\begin{align} \label{eq:r_d} 
\boldsymbol{r}_{\rm ns} &= r_{\rm b} \begin{pmatrix}
\cos{f_{\rm t,o}} \\
\sin{f_{\rm t,o}} \\
0
\end{pmatrix} + r_1 \begin{pmatrix}
\cos(f_{\rm t,i}+\omega)\cos{\Omega}-\cos{\iota}\sin(f_{\rm t,i}+\omega)\sin{\Omega} \\
\cos{\iota}\cos{\Omega}\sin(f_{\rm t,i}+\omega)+\cos(f_{\rm t,i}+\omega)\sin{\Omega} \\
\sin{\iota}\sin(f_{\rm t,i}+\omega)
\end{pmatrix} ,
\end{align}
%
where the first term on the right-hand side is the position vector of the binary barycenter in the triple source frame, with $r_{\rm b} = {a_{\rm o} (1-e_{\rm o}^2)}/({1+e_{\rm o} \cos{f_{\rm t,o}}})$. The second term denotes the position vector of the spinning NS in the binary barycenter frame, with $r_1 = {a_1 (1-e_{\rm i}^2)}/({1+e_{\rm i} \cos{f_{\rm t,i}}})$. The semimajor axis of the NS orbit is $a_1=m_2/(2\pi M/P_{\rm i})^{2/3}$. The true anomaly $f_{\rm t}$ can be expanded in terms of the mean anomaly $M_{\rm a} = 2\pi(t - t_0)/P$ as follows \citep{1988fcmbook},
$\cos{f_{\rm t}} = -e + \frac{1-e^2}{e} \sum_{n=1}^{\infty} 2 J_n(ne) \cos\left({nM_{\rm a}}\right),
\sin {f_{\rm t}} = \sqrt{1 - e^2} \sum_{n=1}^{\infty} \frac{2}{n} \frac{dJ_n(ne)}{de} \sin \left({nM_{\rm a}}\right)$.
Note that $P$ and $e$ refer to the period and eccentricity of either the inner orbit or the outer orbit.
Harmonics truncation is set to $n_{\text{max}} = \left\lfloor 5 {(1 + e)^{1/2}}/{(1 - e)^{3/2}} \right\rfloor$, ensuring that $99\%$ of the signal power is retained \citep{OLeary:2008myb, Mikoczi:2012qy}. 
Here, $\lfloor \cdot \rfloor$ is the floor function. 

Based on the dynamical timescale analysis presented in Sec. \ref{sec-timescales}, the effects of general relativity can be neglected. The outer orbital plane is chosen as the reference plane, the outer orbital elements are assumed to be constant, and the inner orbital elements can be solved via the following secular evolution equations at quadrupole order ($da_{\rm i}/dt=0$) \citep{Will:2017vjc}:
\begin{subequations} \label{eq-secular}
\begin{align} 
\frac{de_{\rm i}}{d\tau} &= 5A \frac{e_{\rm i}(1 - e_{\rm i}^2)^{1/2}}{(1 - e_{\rm o}^2)^{3/2}} \sin^2 \iota \sin\omega \cos\omega , \label{eq-secular1} \\
\frac{d\iota}{d\tau} &= -5A \frac{e_{\rm i}^2 \sin \iota \cos \iota}{(1 - e_{\rm i}^2)^{1/2}(1 - e_{\rm o}^2)^{3/2}}  \sin\omega \cos\omega , \\
\frac{d\Omega}{d\tau} &= -A \frac{\left(1 + 4e_{\rm i}^2 - 5e_{\rm i}^2 \cos^2 \omega \right)}{(1 - e_{\rm i}^2)^{1/2}(1 - e_{\rm o}^2)^{3/2}} \cos \iota , \\  
\frac{d\varpi}{d\tau} &= A \frac{(1 - e_{\rm i}^2)^{1/2}}{(1 - e_{\rm o}^2)^{3/2}} \left[1 - \sin^2 \iota (4 - 5 \cos^2 \omega) \right] , \label{eq-secular4}
\end{align}
\end{subequations}
where a dimensionless time scaled by the inner orbital period $\tau \equiv t/P_{\rm i}$, the coefficient $A \equiv (3\pi/2)m_3/(m_1+m_2)(a_{\rm i}/a_{\rm o})^3$, and the auxiliary variable is defined as ${d\varpi}/{d\tau} \equiv {d\omega}/{d\tau}+{d\Omega}/{d\tau} \cos \iota$.
For simplicity, the effects of the octupole (vanishing for equal mass inner binaries) and hexadecapole order terms \citep{Will:2017vjc}, the spin of the SMBH acting as the tertiary body \citep{Fang:2019hir}, the strong gravity background of Kerr SMBH \citep{Camilloni:2023xvf}, the orbital flips by dark matter dynamical friction \citep{Hu:2024hzu}, and the interactions with surrounding stars and gas \citep{Su:2025} at the Galactic Center are all neglected in our work. The analytical solution based on the equivalent pendulum model, which is convenient for subsequent analysis, is derived in Appendix \ref{sec:analysol}.

The optimal signal-to-noise ratio (SNR) $\rho_{\rm NS}$ for the spinning NS with a monochromatic signal of GW frequency $f$ is given by
\begin{equation}\label{eq:snrns}
\rho_{\rm NS} \approx \sqrt{ \frac{2}{S_n(f)} \int_{0}^{T_{\rm obs}}  h(t)^2 dt } .
\end{equation}
Here, $S_n(f)$ denotes the noise power spectral density and $T_{\rm obs}$ is the observation time. Assuming the 40 km Cosmic Explorer design optimized for the low-frequency band \citep{CE2022slt}, the instrumental noise spectral density depends on frequency and thus varies with NS spin period. We set $T_{\rm obs}=4~{\rm yr}$, the polarization angle $\psi_{\rm p}=\pi/4$, and the angular parameters associated with the detector are adopted as  $\zeta=\pi/2$, $\lambda=0.764$, $\gamma_{\rm o}=1.5$, and $\phi_{\rm r}=\phi_{\rm o}=0$ \citep{Feng:2024ect}.

\begin{figure}[t]
\centering
\includegraphics[scale=0.8]{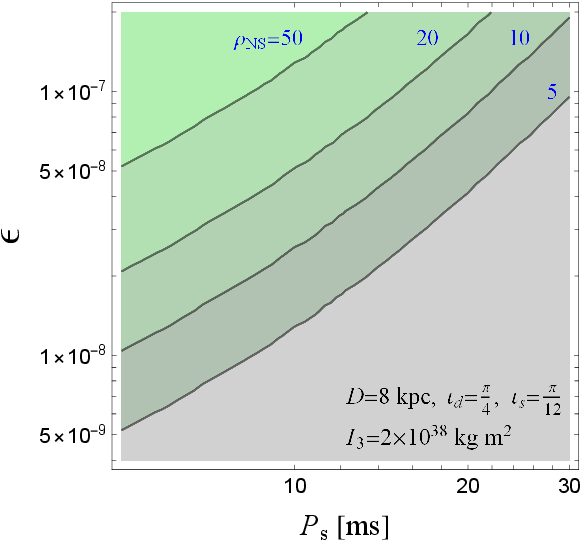}
\caption{SNR contours for continuous GW detection from rapidly spinning NSs at the Galactic Center, shown in the spin period–equatorial ellipticity parameter space ($P_{\rm s}$, $\epsilon$) for a 4-year observation of Cosmic Explorer. The green shaded region indicates the detectable parameter space (assuming an SNR threshold of $\rho_{\rm NS} = 5$). NSs spinning at $P_{\rm s}=10$ ms are detectable with ellipticities as small as $\epsilon \sim 10^{-8}$. }
\label{fig:SNRforNS}
\end{figure}

Figure \ref{fig:SNRforNS} presents SNR contours for rapidly spinning NSs at the Galactic Center in the parameter space of spin period ($P_{\rm s}$) and equatorial ellipticity ($\epsilon$) for a 4-year observation of Cosmic Explorer, calculated for a fixed distance of $D=8~{\rm kpc}$, NS moment of inertia $I_3=2\times 10^{38}~{\rm kg~m^2}$, and specific angular parameters (e.g., $\iota_{s}=\pi/12$ and $\iota_{d}=\pi/4$). The contours are consistent with the scaling of continuous GW amplitude with $\epsilon \, \Omega_{\rm r}^2$ in Eq.~(\ref{eq-2polarizations}). 
The gray region ($\rho_{\rm NS}<5$) represents parameter space below the detection threshold, while the green regions ($\rho_{\rm NS} \ge 10$, up to $\rho_{\rm NS}=50$) denote robustly detectable signals. For example, an NS with a spin period of $10~\rm{ms}$ can be detected down to an ellipticity of $\epsilon \approx 10^{-8}$, whereas an NS with $P_{\rm s}=30~\rm{ms}$ requires $\epsilon \gtrsim 10^{-7}$ to reach the $\rho_{\rm NS}=5$ detection threshold. These results highlight the interplay between NS spin and ellipticity in determining detectability and demonstrate that rapidly spinning NSs with even modest ellipticities are promising targets for continuous-wave detection with Cosmic Explorer in the Galactic Center.

\section{ZLK effect-induced dual-line gravitational radiation}
\label{sec:zlkinduced}

The SNR for an inspiraling NS--NS system to be detected by space-borne GW observatories, characterized by an effective noise power spectral density $S_n(f)$, can be approximated as follows (see \citealt{Wang:2020jsx} and references therein):
\begin{equation}\label{eq:snrbinary}
\rho_{\rm NS\text{–}NS}(e_{\rm i},T_{\rm obs}) \approx \frac{32\sqrt{2}}{5} \frac{m_1 m_2}{a_{\rm i} D} \sqrt{ 0.886~T_{\rm obs} \sum_{n=1}^{n_{\max}} \frac{g(n,e_{\rm i})}{n^2 S_n(f_n)}}  ,
\end{equation} 
where $f_n=n/P_{\rm i}$ denotes the $n$-th orbital frequency harmonic, and $g(n,e_{\rm i})$ is the radiation power factor for the $n$-th harmonic defined via the Bessel functions of the first kind \citep{Peters:1963ux}.

\begin{figure*}[t]
\plottwo{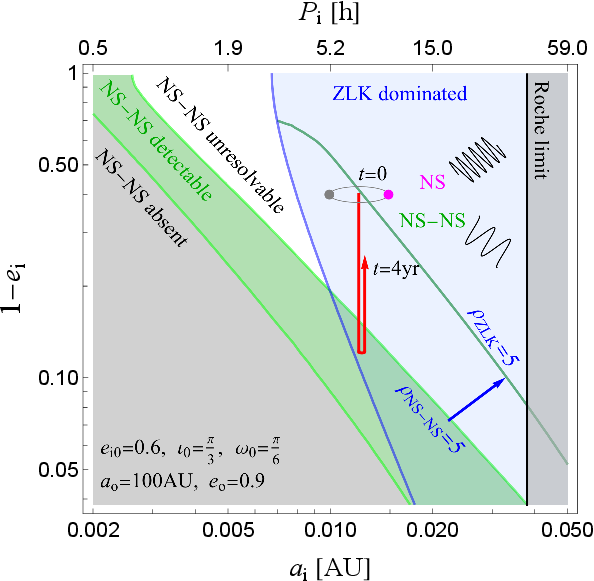}{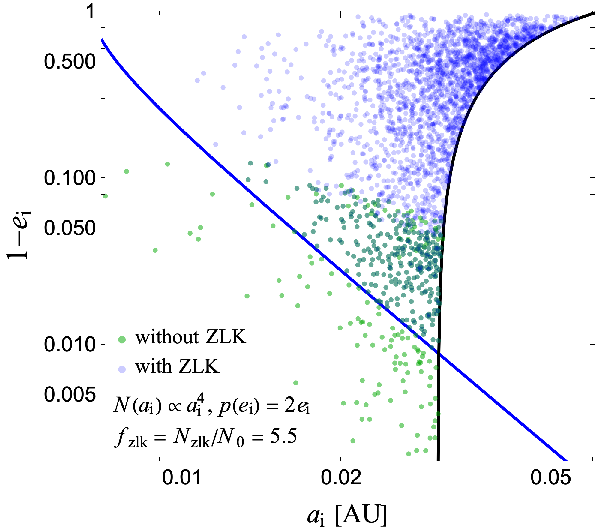} 
\caption{Dual-line gravitational radiation excitation through the ZLK effect for the Galactic Center NS--NS systems. Left panel: Green contours show LISA-detectable systems without the ZLK effect, bounded by $\rho_{\rm NS\text{–}NS} = 5$ (right) and $\rho_{\rm NS\text{–}NS} = 40$ (left) in the $a_{\rm i}$–$(1-e_{\rm i})$ parameter space. The light blue region indicates ZLK dominance, exemplified by a system with $e_{\rm i0} = 0.6$ and $P_{\rm i} = 7$ h. The magenta point represents a rapidly spinning NS component detectable by Cosmic Explorer (Fig. \ref{fig:SNRforNS}). Without ZLK, this system has $\rho_{\rm NS\text{–}NS} = 0.16$ (undetectable), but ZLK-driven eccentricity oscillations boost the SNR to 5.5 over 4 years. The dark green contour shows the ZLK-modified SNR threshold $\rho_{\rm ZLK} = 5$, which extends into regions of larger orbital separation and lower eccentricity than the standard detection boundary. Right panel: Monte Carlo sampling of inner binary orbital elements, followed by ZLK-induced secular evolution calculations, indicates that this parameter‑space expansion increases the number of detectable sources by $\sim 5$-fold (or nearly an order of magnitude for the uniform eccentricity distribution; see Fig. \ref{fig:fzlkuniform} in Appendix \ref{sec:calfzlk}). ZLK oscillations therefore promote wide binaries from unresolvable backgrounds into LISA's detection window, substantially improving dual‑line prospects and establishing the ZLK effect as a viable formation channel for Galactic Center dual-line systems.}
\label{fig:duallineexc}
\end{figure*}

For isolated NS--NS systems without the ZLK effect [neglecting the variation of $e_{\rm i}$ in Eq. (\ref{eq:snrbinary})], the $a_{\rm i}$–$(1-e_{\rm i})$ parameter space for LISA detectability naturally divides into three distinct regions in Fig. \ref{fig:duallineexc}: the upper white region (unresolvable), the green diagonal band (detectable), and the lower gray region (absent). The physical origin of this division reflects GW-driven evolution and source population statistics. Gravitational radiation causes orbital decay, reducing both semimajor axis and eccentricity \citep{Peters:1964zz}, which concentrates sources toward larger $a_{\rm i}$ and higher $e_{\rm i}$ values \citep{Feng:2023fez, Feng:2024ulg, Kyutoku:2016ppx}.
This evolutionary bias creates three regions: (1) The white region contains wide-orbit systems that emit GWs below the detection SNR threshold of $\rho_{\rm NS\text{–}NS}=5$, forming an unresolved background; (2) The green band represents the optimal detection window where binary inspirals produce sufficiently strong signals, e.g., $\rho_{\rm NS\text{–}NS}=5\text{–}40$ as found in simulations of \citet{Wang:2020jsx}; (3) The gray region is depleted of sources because tight binaries merge rapidly, spending minimal time in this high frequency regime—consistent with simulations of \citet{Wang:2020jsx} showing no systems exceeding SNR $\sim 40$.

This stable three-region structure changes significantly when NS--NS systems are embedded in hierarchical triple systems with Sgr A$^*$. The ZLK effect introduces periodic modulations in orbital eccentricity, allowing systems to migrate between detection regimes over evolutionary timescales.
ZLK oscillations dominate over general relativistic precession in the light blue region of Fig. \ref{fig:duallineexc}. Neglecting the $e_{\rm i}$ dependence and order‑unity factors, this is approximately equivalent to $P_{\rm ZLK} < P_{\rm 1PN}$, which corresponds to the hierarchical triple criterion \citep{Blaes:2002cs}:
\begin{equation}
\frac{a_{\rm o}^3}{a_{\rm i}^3} < \frac{3 m_3 a_{\rm i} \left(1 - e_{\rm i}^2\right)^{3/2}}{4 (m_1 + m_2)^2 \left(1 - e_{\rm o}^2\right)^{3/2}} .
\end{equation}
%
In ZLK-dominated systems, orbital eccentricity varies significantly over the observation period, requiring a time-dependent analysis. We discretize the total observation time $T_{\rm obs}$ into $N_{\rm seg}$ segments of duration $\Delta T$, treating the eccentricity as constant ($e_{{\rm i},j}$) within each interval. The integrated SNR is computed by summing the squared SNR contributions from each segment:
\begin{equation}\label{eq:snrZLK}
\rho_{\rm ZLK}  \approx \sqrt{ \sum_{j=1}^{N_{\rm seg}} \rho^2_{\rm NS\text{–}NS}(e_{{\rm i},j},\Delta T)}  .
\end{equation} 

Figure \ref{fig:duallineexc} demonstrates the dual-line gravitational radiation excitation through the ZLK effect. 
In the ZLK-dominated regime, we examine a representative system with initial parameters matching the most probable system in \citet{Wang:2020jsx}: $e_{\rm i0} = 0.6$, $P_{\rm i} = 7~\rm{h}$ (binary labeled at $t=0$), and a rapidly spinning NS (magenta dot) detectable by Cosmic Explorer (Fig. \ref{fig:SNRforNS}). 
We take the outer orbital plane as the reference frame, fixing its inclination and longitude of ascending node to 0. The following initial values for the inner binary are adopted here exclusively for a specific illustrative example system (see the left panel of Fig. \ref{fig:duallineexc}): inclination $\iota_0 = \pi/3$, longitude of pericenter $\omega_0 = \pi/6$, and longitude of ascending node $\Omega_0 = \pi/8$. The secular evolution of inner binary orbital elements is obtained via direct numerical integration of Eqs. (\ref{eq-secular1})--(\ref{eq-secular4}).
The ZLK enhancement is dramatic: without ZLK effects, this system yields $\rho_{\rm NS\text{–}NS} = 0.16$ [Eq. (\ref{eq:snrbinary})], remaining undetectable by LISA. However, ZLK-driven eccentricity oscillations over 4 years boost the SNR to $\rho_{\rm ZLK} = 5.5$ [Eq. (\ref{eq:snrZLK})], crossing the detection threshold.
We find that the ZLK‑modified SNR contour ($\rho_{\rm ZLK}=5$, dark green) extends to larger orbital separations and lower eccentricities, substantially increasing the detectable population. 

To quantify this enhancement (see the right panel of Fig. \ref{fig:duallineexc}), we perform Monte Carlo sampling of the inner orbital parameters: the semimajor axis $a_{\rm i}$ is drawn from a distribution $\propto a_{\rm i}^4$ \citep{Feng:2024ulg}) and the eccentricity from a thermal law $p(e_{\rm i})=2e_{\rm i}$ \citep{Heggie:1975rcz}. The angular orbital elements are uniformly sampled: $\cos{\iota_0}\in[-1,1]$, $\omega_0\in[0,2\pi]$, $\Omega_0\in[0,2\pi]$. For each realization, we numerically integrate Eqs. (\ref{eq-secular1})--(\ref{eq-secular4}) to compute the orbital evolution. We count the number of sources $N_{\rm zlk}$ that exceed the detection threshold of 5 and that lie within the region bounded by the ZLK-dominated boundary and the Roche‑limit defined below. We account for the inclination dependence of the SNR using the factor $\{(5/4)[\cos^2\iota+(1+\cos^2\iota)^2/4]\}^{1/2}$ \citep{Robson:2018ifk}, and neglect the impact of $\omega$ and $\Omega$ on the SNR. A more comprehensive treatment of these angular dependencies will require refined waveform modeling in future work. For comparison, we count the number of detectable sources $N_0$ assuming fixed, unevolving initial orbital elements (i.e., neglecting ZLK oscillations). We then define the ZLK-induced detection enhancement factor as $f_{\rm zlk} \equiv N_{\rm zlk}/N_0$. This procedure indicates that allowing for ZLK‑induced excursions into wider separations and lower eccentricities yields a $\sim 5$-fold (nearly an order of magnitude) enhancement in the detection count for the thermal (uniform) eccentricity distribution. A detailed calculation procedure is provided in Appendix \ref{sec:calfzlk}, where we validate the robustness of the enhancement factor by repeating the analysis for all combinations of two eccentricity distributions (thermal and uniform) and two SNR thresholds (5 and 8).

As an order‑of‑magnitude feasibility estimate, we then write
\begin{equation}
N_{\rm d} \sim N_{\rm b} \times f_{\rm 100au} \times f_{\rm ns} \times f_{\rm zlk} ,
\end{equation}
where $N_{\rm b} \sim 0.6\text{–}6$ denotes the number of LISA-detectable NS binaries \footnote{This conservative estimate aligns with the simple scaling of the Galactic NS-hosting binary population to the inner 0.1 pc of Sgr A* from \citet{Tang:2024lst}, which sets its lower bound and illustrates the Galactic Center dynamical boosts to compact binary formation.} assuming constant eccentricity during the observation period [Eq. (\ref{eq:snrbinary})] \citep{Wang:2020jsx}, $f_{\rm 100au} \sim 0.1$ represents the fraction of binaries within 100 AU of Sgr A$^*$ (assuming log-uniform outer orbits to 0.1 pc), and $f_{\rm ns} \sim 0.2\text{–}0.6$ is the fraction of rapidly spinning NS components in LISA-detectable dual-line binaries for Cosmic Explorer if no ZLK was present \citep{Feng:2025jnx}. Using the Monte Carlo-derived ZLK enhancement factor ($f_{\rm zlk} \sim 5\text{–}10$) raises the expected dual‑line yield from a conservative estimate of 0.01--0.4 (neglecting ZLK orbital evolution) to 0.05--4 over a 4-year observation period. These order‑of‑magnitude results indicate that ZLK dynamics constitute a viable channel for producing Galactic Center dual‑line systems.

Note that the binary considered in the above example satisfies the stability criterion requiring that the inner binary not cross the Roche limit of the Sgr A$^*$ at its pericenter \citep{Hoang:2019kye, Naoz_2014}, namely ${a_{\rm i}}/{a_{\rm o}} < \left[(m_1 + m_2)/({3 \, m_3})\right]^{1/3} ({1 - e_{\rm o}})/({1 + e_{\rm i}})$.
Additionally, it satisfies the condition for the applicability of the orbit-averaged approximation, valid for a stationary outer perturber and instantaneous quadrupole torque \citep{Antonini:2013tea}, as $\sqrt{1-e_{\rm i}} > 5\pi {m_3}/({m_1 + m_2})~{a_{\rm i}}^3/[{a_{\rm o}(1 - e_{\rm o})}] ^3$.
To facilitate the visualization of the eccentricity evolution path, we intentionally introduce a small change in the inner orbital semimajor axis $a_{\rm i}$ over a 4-year observation period, manifested as the non-overlapping of the two red vertical lines in Fig. \ref{fig:duallineexc}.
Although 1PN precession is enhanced by a factor of $(1-e_{\rm i}^2)^{-1}$ [see Eq. (\ref{eq-1pn})] during eccentricity excitation, it remains insufficient to suppress ZLK oscillations in the light blue region where most Galactic Center sources reside.

\section{Discussion and Conclusions}\label{sec-conclusion}

In this work, we demonstrate that the ZLK effect in Galactic Center hierarchical triples can dramatically enhance the prospects for simultaneous multi-band dual-line GW detection. Our dynamical analysis reveals that ZLK oscillations dominate over general relativistic precession for wide NS--NS binaries, enabling significant orbital modulations over multi-year timescales. 
This mechanism operates through complementary pathways: ZLK-driven eccentricity oscillations promote wide systems from below LISA sensitivity into the detectable regime, while rapidly spinning NSs in these binaries simultaneously emit continuous GWs accessible to Cosmic Explorer. The resulting dual-line detectability represents a substantial enhancement over isolated binary evolution.

Quantitatively, we find that ZLK-induced orbital evolution extends the detectable parameter space to larger separations and lower eccentricities, yielding a $\sim 5$-fold enhancement for the fiducial thermal eccentricity distribution and $\sim 10$-fold for the uniform eccentricity distribution. We estimate an expected yield of 0.05--4 dual‑line sources per 4‑year mission, compared to a conservative estimate of 0.01--0.4 without ZLK evolution, transforming the observational outlook from challenging to more feasible.

We emphasize that our adopted SNR thresholds (5 and 8) are minimum values for source detectability, not requirements for high-precision NS parameter measurements or independent distance calibration. The latter typically requires $\sim 30$ for isolated low-eccentricity NS binaries to resolve the degeneracy between NS structural parameters and distance \citep{Feng:2023fez, Feng:2025jnx}. For the ZLK-modulated systems considered here, sources satisfying the chirp detectability condition (for distance measurement) have total SNRs of 10--50, with a fraction reaching this optimistic $\sim 30$ threshold. While time-varying eccentricity introduces additional waveform parameters, ZLK-induced amplitude and phase modulations may mitigate some degeneracies, and richer harmonic structure in higher-eccentricity signals provides complementary constraints. Whether meaningful dual-line NS parameter measurements can be achieved below 30, and if so how far below, remains an open question, dependent on future advances in waveform modeling and data analysis techniques.

The successful detection of ZLK-induced dual-line emission relies on the synergy of space-borne and ground-based detectors: LISA first detects the low-frequency inspiral signal to obtain precise sky localization, orbital parameters, and system geometry of the NS binary, and Cosmic Explorer or Einstein Telescope then performs a directed search for the high-frequency spinning NS signal within the constrained parameter space (Fig. \ref{fig:SNRforNS}). This two-step strategy abandons blind all-sky searches for spinning NSs, and the waveform template of spinning NSs in ZLK-modulated triples presented in this work provides the key input for developing dedicated data analysis pipelines for such dual-line sources. 

Our estimates rely on simplified population simulations and assume favorable system configurations. Future studies should incorporate more realistic stellar-evolution models, detailed ZLK modeling that includes high-order effects, and comprehensive population synthesis to refine these predictions.

ZLK oscillations thus provide a promising formation channel for Galactic Center dual-line sources, offering significant scientific opportunities beyond detection enhancement. Simultaneous multi-band observations would place tight constraints on NS equations of state while encoding Galactic Center dynamics, establishing a valuable pathway for dual-line GW astronomy that bridges compact object physics and Galactic structure.

\begin{acknowledgments}

We thank the anonymous referee for invaluable comments that have significantly improved this manuscript.
We thank Smadar Naoz, Xian Chen, Yan Wang, Haoran Di, Hao Wang, Zhao Li, and Jiang-Chuan Yu for helpful discussions.    
W.-F.F. is supported by the China Postdoctoral Science Foundation under Grant No. 2025M783222, and the National Natural Science Foundation of China under Grant No. 12447109.
B.L. acknowledges support from the National Natural Science Foundation of China (Grant No. 12433008) and National Key Research and Development Program of China (No. 2023YFB3002502).
L.S. acknowledges support from the Beijing Natural Science Foundation (1242018), the National Natural Science Foundation of China (12573042), the National SKA Program of China (2020SKA0120300), and the Max Planck Partner Group Program funded by the Max Planck Society.

\end{acknowledgments}

\appendix

\section{Analytical solution for eccentricity and inclination} \label{sec:analysol}

The quadrupole ZLK mechanism for a test particle exhibits behavior analogous to that of a mechanical pendulum \citep{Basha2025}, which is described by the following equation:
\begin{equation}
\ddot{\theta}_{\rm K} + \omega_{\rm K}^2 \sin\theta_{\rm K} =0,
\end{equation}
where $\theta_{\rm K}$ denotes the pendulum angle, and $\omega_{\rm K}$ is a constant corresponding to the angular frequency in small-amplitude oscillations. In the context of two conserved quantities of motion: $j_z=\left(1-e_{\rm i}^2 \right)^{1/2} \cos \iota$ and $C_{\rm K}=e_{\rm i}^2 \left(1-5\sin^2\iota \sin^2 \omega /2\right)$, this analogy is explicitly expressed as:
\begin{subequations}
\begin{align}
\dot{\theta}_{\rm K} &= \frac{3}{2}\sqrt{15}\, e_{\rm i} \sin \iota \sin\omega \,, \\
\omega_{\rm K}^2 &= \frac{9}{8}\left[\left(3 - 5 j_z^2 - 2 C_{\rm K} \right)^2 + 24 C_{\rm K}\right]^{1/2} \,.
\end{align}
\end{subequations}
The normalized energy of the pendulum (a dimensionless constant of motion) is given by:
\begin{equation}
E_{\rm K} = \frac{\dot{\theta}_{\rm K}^2}{2 \omega_{\rm K}^2} + 1 - \cos\theta_{\rm K} = 1 + \frac{9}{8 \omega_{\rm K}^2} \left( 3 - 5 j_z^2 - 8 C_{\rm K} \right) \,.
\end{equation}

The exact solution for the angular displacement $\theta_{\rm K}(t)$ under the initial conditions, $\theta_{\rm K}(0) = \theta_{\rm K0}$ and $\dot{\theta}_{\rm K}(0) = \omega_{\rm K0}$, is given by:
\begin{equation}
\theta_{\rm K}(t) = 2 \arcsin\left[ k \cdot \operatorname{sn} \left( \pm \omega_{\rm K} t + \phi_{\rm K0}, k \right) \right] \,,
\end{equation}
where $\operatorname{sn}(u, k)$ denotes the Jacobian elliptic sine function. The modulus $k$ and initial phase constant $\phi_{\rm K0}$ of this elliptic function are defined as:
\begin{equation}
k = \sin \left( \frac{\theta_{\rm Kmax}}{2} \right)\,, \quad \phi_{\rm K0} = \operatorname{sn}^{-1}\left( \frac{\sin(\theta_{\rm K0} / 2)}{k}, k \right) \,,
\end{equation}
with $\theta_{\rm Kmax}$ representing the maximum angular displacement of the pendulum-like motion. For pendulum libration ($C_{\rm K}>0$), $\theta_{\rm Kmax}=\arccos(1-E_{\rm K})$.
The sign “$\pm$” in the argument of $\operatorname{sn}(u, k)$ is determined by the sign of $\omega_{\rm K0}$.

The analytical solution for the inner eccentricity reads: 
\begin{equation}\label{eq-analyei}
e_{\rm i} = \left[l_{\rm K}E_{\rm K}+C_{\rm K}-l_{\rm K}(1-\cos{\theta_{\rm K}})\right]^{1/2} \,,
\end{equation}
where $l_{\rm K}=4\omega_{\rm K}^2/27$ denotes the equivalent length of the pendulum (a derived constant of the system).
Correspondingly, the analytical solution for the inclination is given by:
\begin{equation}
\iota = \arccos\left[{j_z\left(1-e_{\rm i}^2 \right)^{-1/2}}\right] \,.
\end{equation}

\section{Calculation of the ZLK enhancement factor} \label{sec:calfzlk}

This appendix provides a detailed description of the Monte Carlo simulation procedure used to compute the ZLK-induced detection enhancement factor $f_{\rm zlk} \equiv N_{\rm zlk}/N_0$, where $N_{\rm zlk}$ and $N_0$ are the numbers of detectable sources in the ZLK-dominated and reference regions, respectively. We also present the robustness tests conducted to verify the reliability of the derived enhancement factor.
The custom code implementing this Monte Carlo simulation and $f_{\rm zlk}$ calculation is publicly available \citep{Feng2026ZLKCode}.

\textbf{Parameter Space and Sampling Distributions.} We consider the orbital parameter space spanned by the semimajor axis $a_{\rm i}$ and the eccentricity $e_{\rm i}$ (plotted as $1-e_{\rm i}$ in the main text figure). The sampling range for the semimajor axis is fixed to $a_{\rm i} \in (a_{\rm min}, a_{\rm max})$, where $a_{\rm min}=0.007~{\rm AU}$ (the lower boundary of interest) and $a_{\rm max}=a_{\rm R}=0.06~{\rm AU}$ (the Roche limit for zero eccentricity). Within this parameter space, the Roche limit is defined by the curve satisfying ${a_{\rm i}}/{a_{\rm o}} < \left[(m_1 + m_2)/({3 \, m_3})\right]^{1/3} ({1 - e_{\rm o}})/({1 + e_{\rm i}})$ \citep{Hoang:2019kye, Naoz_2014}. Eccentric collisions do not occur in our simulations, as $a_{\rm i}(1-e_{\rm i})$ is significantly larger than the radii of the NS. For all sampling schemes, we generate a total of $10^5$ random sample points to ensure high statistical precision and convergence of the source counts $N_{\rm zlk}$ and $N_0$.

The semimajor axis $a_{\rm i}$ is sampled with a probability distribution $\propto a_{\rm i}^4$, physically motivated by the binary distribution \citep{Feng:2024ulg}. This sampling is implemented via inverse transform sampling: a uniform random deviate $u_a \in (0, 1)$ is mapped to $a_{\rm i}$ using the formula $a_{\rm i} = \left( u_a \left(a_{\rm i, max}^5 - a_{\rm i, min}^5\right) + a_{\rm i, min}^5 \right)^{1/5}$.
The angular orbital elements are sampled from uniform distributions to account for the random orientation of the inner binary: the cosine of the initial inclination $\cos\iota_0$ is uniformly sampled over [-1,1], while the initial argument of pericenter $\omega_0$ and initial longitude of ascending node $\Omega_0$ are each independently uniformly sampled over $[0,2\pi]$.
For the initial eccentricity $e_{\rm i0}$, we employ two distinct sampling schemes:
\begin{enumerate}
    \item[(i)] \textit{Thermal Eccentricity Sampling.} We adopt the thermal eccentricity distribution with a probability density function (PDF) $p(e_{\rm i}) = 2e_{\rm i}$ for $e_{\rm i} \in (0, 1)$ \citep{Heggie:1975rcz}. Since our parameter space is presented in terms of $y \equiv 1-e_{\rm i}$, we perform sampling in this transformed variable. First, we compute the normalization constant for the PDF: $C = (1-y_{\rm min})^2 - (1-y_{\rm max})^2$. We then generate a uniform deviate $u_y \in (0, 1)$ and apply the inverse transform sampling formula to obtain $y = 1 - \sqrt{(1-y_{\rm min})^2 - u_y \cdot C}$.   
    \item[(ii)] \textit{Uniform Eccentricity Sampling.} The eccentricity $e_{\rm i}$ (and thus $1-e_{\rm i}$) is sampled uniformly over its full physical range $e_{\rm i} \in (0, 1)$. A uniform random deviate $u_e \in (0, 1)$ is directly mapped to $1-e_{\rm i} = u_e$.
\end{enumerate}

\textbf{Orbital Evolution and SNR Calculation.} For each Monte Carlo realization, we compute the secular orbital evolution of the inner binary by numerically integrating the ZLK secular equations [Eqs. (\ref{eq-secular1})--(\ref{eq-secular4})]. 
The total observation time $T_{\rm obs}$ is divided into $N_{\rm seg}=20$ equal segments of duration $\Delta T = T_{\rm obs}/N_{\rm seg}$. Within each segment, the evolving eccentricity $e_{{\rm i},j}$ and inclination $\iota_j$ are treated as constant. For simplicity, if the eccentricity exceeds the ZLK-dominated threshold (blue curve in Fig.~\ref{fig:duallineexc}) at any point during the secular evolution, it is held constant for all subsequent evolutionary segments.
The SNR for each source is calculated by incorporating the inclination dependence of the waveforms, using the correction factor $\{(5/4)[\cos^2\iota + (1 + \cos^2\iota)^2/4]\}^{1/2}$ \citep{Robson:2018ifk}. The total SNR accounting for ZLK-induced orbital evolution, denoted $\rho_{\rm ZLK}$, is given by: 
\begin{equation}\label{eq:snrZLKwithInc}
\rho_{\rm ZLK}  \approx \sqrt{ \sum_{j=1}^{N_{\rm seg}} \frac{5}{4}\left[\cos^2\iota_j+\frac{(1+\cos^2\iota_j)^2}{4} \right] \rho^2_{\rm NS\text{–}NS}(e_{{\rm i},j}, \Delta T)}  .
\end{equation} 
As noted in the main text, we neglect the impact of $\omega$ and $\Omega$ on the SNR; a more comprehensive treatment of these angular dependencies will require refined waveform modeling in future work.

\textbf{Detectable Source Counting and Visualization.}
We adopt two values for the SNR detection threshold throughout our Monte Carlo analyses: a fiducial threshold of $\rho=5$, from population synthesis simulations \citep{Wang:2020jsx}, for our baseline calculation (as used in the main text), and a more stringent threshold of $\rho=8$, from simulated LISA observations \citep{Knee:2024mst}, to test the robustness of our results. We define two sets of detectable source counts to compute $f_{\rm zlk} \equiv N_{\rm zlk}/N_0$:
\begin{enumerate}
    \item[(i)] ZLK-included count $N_{\rm zlk}$: Sources that satisfy two criteria: (1) the SNR exceeds the detection threshold (fiducially $\rho = 5$; $\rho = 8$ for robustness tests), and (2) the orbit lies within the ZLK-dominated region and inside the Roche limit. 
    \item[(ii)] Reference count $N_0$: Sources that assume fixed, unevolving initial orbital elements (i.e., neglecting ZLK-induced orbital evolution). The detection criteria require an SNR above the threshold and that the system lies within the Roche limit.
\end{enumerate}

\begin{figure*}[htbp] 
\centering
\includegraphics[scale=0.76]{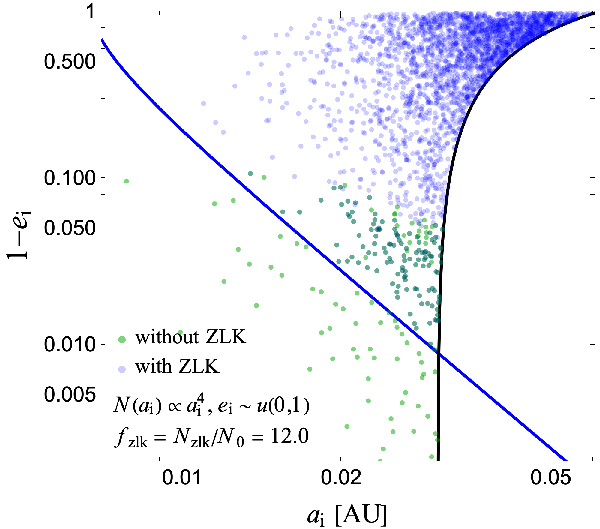}
\caption{As in the right panel of Fig. \ref{fig:duallineexc}, but for uniform eccentricity sampling, yielding $f_{\rm zlk}=12$. }
\label{fig:fzlkuniform} 
\end{figure*}

\begin{figure*}[htbp]
\plottwo{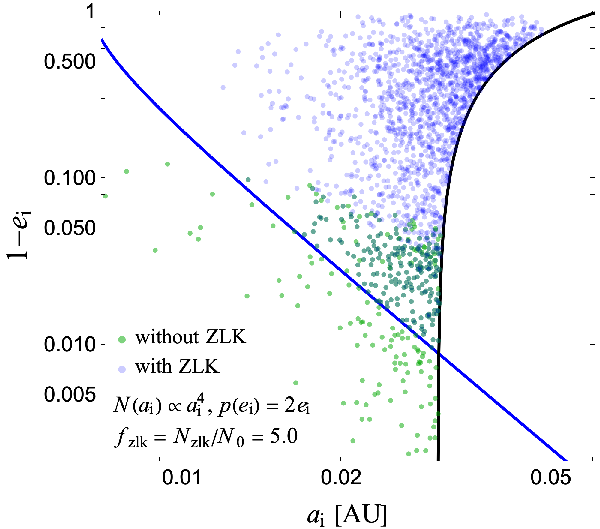}{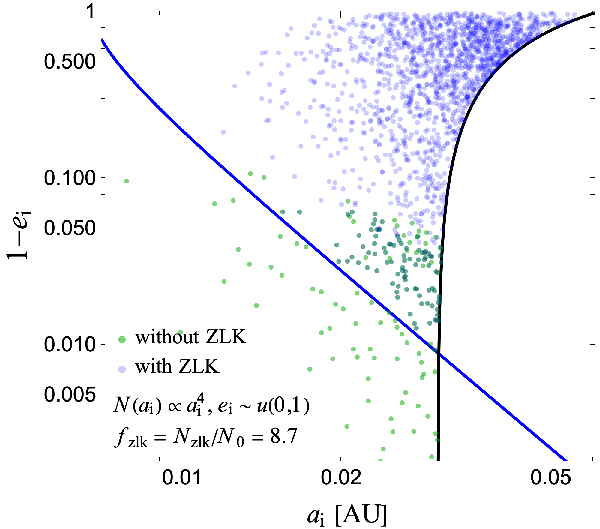} 
\caption{As in Fig. \ref{fig:fzlkuniform}, but with a binary SNR threshold of 8, yielding $f_{\rm zlk}=5.0$ (thermal) and $f_{\rm zlk}=8.7$ (uniform) for the two eccentricity sampling schemes.}
\label{fig:fzlksnr8} 
\end{figure*}

We show the initial orbital parameters $(a_{\rm i}, 1-e_{\rm i})$ of detectable sources as individual scatter points in Figs.~\ref{fig:duallineexc}, \ref{fig:fzlkuniform}, and \ref{fig:fzlksnr8}, where each point represents one Monte Carlo realization that meets the detectability criteria described earlier.

\textbf{Robustness Tests.} To validate the robustness of the derived enhancement factor, we perform three independent Monte Carlo re-analyses, covering all combinations of two eccentricity distributions and two SNR thresholds:
\begin{enumerate}
    \item[(i)] Uniform eccentricity distribution + SNR threshold $\rho = 5$;
    \item[(ii)] Thermal eccentricity distribution (fiducial) + SNR threshold $\rho = 8$;
    \item[(iii)] Uniform eccentricity distribution + SNR threshold $\rho = 8$.
\end{enumerate}

All re-analyses use the same parameter sampling range, integration settings, and detection criteria as the fiducial calculation (thermal eccentricity + $\rho = 5$). The larger enhancement factor for the uniform eccentricity distribution arises from the shape of the sampling distribution. Increasing the binary SNR threshold from 5 to 8 causes a mild decrease in $f_{\rm zlk}$. This small variation does not alter our core conclusion: ZLK-driven eccentricity oscillations substantially expand the detectable counts of Galactic Center dual-line GW sources.

%


\bibliography{reference}{}
\bibliographystyle{aasjournalv7}



\end{document}